\documentclass[aps,prl,twocolumn,groupedaddress,showpacs]{revtex4}
\usepackage{epsfig}

\begin{document}


\title{Atomic scale lattice distortions 
and domain wall profiles}

\author{K. H. Ahn, T. Lookman, A. Saxena, and A. R. Bishop} 
\affiliation{Theoretical Division, Los Alamos National Laboratory, 
Los Alamos, New Mexico 87545}

\begin{abstract}
We present an atomic scale theory of lattice distortions
using strain related variables and their constraint equations.
Our approach connects 
constrained {\it atomic length} scale variations
to {\it continuum} elasticity 
and describes elasticity at several length scales.
We apply the approach to a  two-dimensional square lattice 
with a monatomic basis,
and find the elastic deformations and hierarchical atomic relaxations
in the vicinity of a  domain wall between two different 
homogeneous strain states.  
We clarify the microscopic origin of gradient
terms, some of which are included phenomenologically in 
Ginzburg-Landau theory,
by showing that they are anisotropic.
\end{abstract}


\maketitle

\newpage
An understanding of atomic scale lattice distortions is
essential for correctly describing the elastic energies of
nano-structured materials. 
New generations of experimental tools
to probe individual atoms 
and local environments~\cite{Stemmer95}, 
and the growing interest in complex
functional materials, in which local lattice distortions 
are coupled to 
electronic, magnetic, and chemical
degrees of freedom, further emphasize
the need for a consistent theoretical framework to describe
strain-based materials. 
For example, in perovskite manganites 
the change in oxygen ion displacement 
at each site 
is associated with the charge and orbital ordering 
states~\cite{Millis96.Radaelli97}.
An atomic scale description of the interface or  domain wall
between two different homogeneous states is thus a first step towards predicting 
functionality located at the domain walls.

Strain variables (rather than displacement) with constraints
have been recently shown to have advantages
for describing the long wavelength lattice
distortions observed in, for example, martensitic materials
and, more generally, solid-solid phase 
transformations~\cite{Shenoy99}.
The {\it anisotropic long-range} interaction in the order parameter 
strain fields
drives the formation of a rich landscape of 
multiscale elastic textures.
The aim of this work is to formulate a microscopic
description of elasticity and demonstrate 
the relationship with and
differences from long-wavelength continuum theory. 
We introduce
appropriate inter-cell and intra-cell distortion modes and show how the 
form of the elastic energy recovers the correct phonon spectra.
The discreteness of the lattice, choice of modes and constraints
among them give rise to an anisotropic
gradient expansion for the elastic energy.
This leads to elastic domain wall solutions
that are different from those predicted using continuum theory; 
we obtain $0^o$ and $90^o$ `staircase' domain walls 
for sufficiently small bulk modulus (or `soft') materials, in addition to the 
$45^o$ or $135^o$ walls  predicted from continuum theory for `hard'
materials.  

Models based on displacement variables
with pair-potentials, 
such as Born-von K\'{a}rm\'{a}n models~\cite{Kittel},
have been widely used to incorporate `microscopic elasticity'.
However,
essentially because distortion implies strain, 
the physical insight for atomic scale elasticity
will reveal itself in the language of strain-related variables 
presented here.
Moreover, our work is quite distinct from recent efforts to describe
elasticity of nanometer-sized 
objects~\cite{Lu97.Hernandez98.Segall00}. 
The interest there 
is to describe {\it long wavelength} strains in a given dimension with
other dimensions maintained at nanoscales, such as ultra-thin long
nanowires. Our emphasis is to describe 
{\it atomic scale distortions}, irrespective of 
whether the region of interest 
is in bulk or nano-sized objects.
Our approach describes elastic deformation 
in terms of {\it intra}-cell modes or ``shuffles'' of atoms, which
are essential in
describing short wavelength lattice distortions,
and distortion of unit cells,
instead of adopting coarse graining 
approximations~\cite{Goldhirsch02}.

\begin{figure}
\leavevmode
\epsfxsize8.5cm\epsfbox{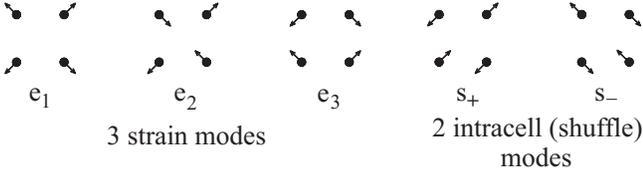}
\caption{\label{fig:squaremode} Normal distortion modes for 
a square object of four atoms in
2D. 
}
\end{figure}

We illustrate our ideas in detail 
for the simplest case, namely a
square lattice in two-dimensional (2D) space
with a monatomic basis.
We find that the most convenient strain-related
variables for atomic scale distortions are  
the normal distortion modes
(more precisely, symmetry coordinates)
of an elementary
square object of four atoms,
as shown in Fig.~\ref{fig:squaremode}.
Because of the number of atoms in this object and the dimensionality,
eight normal modes exist.
The rigid rotation and the two rigid translations
(not shown in Fig.~\ref{fig:squaremode})
cost no elastic energy, and therefore, are not distortion modes.
The first  
three distortion modes in Fig.~\ref{fig:squaremode} 
correspond to the usual dilatation ($e_1$), 
shear ($e_2$), and
deviatoric ($e_3$) strains of the continuum elasticity theory
for a square lattice~\cite{Shenoy99}.
The next two degenerate modes in Fig.~\ref{fig:squaremode}, 
$s_+$ and $s_-$,
correspond to
the ``intracell'' or ``shuffle''
modes of the square lattice~\cite{Barsch84}, 
which are  absent in continuum elasticity theory.
Our approach uses these five distortion variables
defined for each plaquette of four atoms
at $\vec{i}$, $\vec{i}+(10)$, $\vec{i}+(11)$, and $\vec{i}+(01)$,
where $\vec{i}$ represents the coordinate of the lattice points,
to describe the elastic energy~\cite{definition}. 

Since the five variables 
are derived from
two displacement variables for each lattice site,
they are related by three
constraint equations.
By representing $e_1$, $e_2$, $e_3$, $s_+$, and $s_-$
in terms of displacement variables $d^x$ and $d^y$ in $k$ (wavevector) space
and eliminating $d^x$ and $d^y$, 
the constraint equations
are obtained.
One of them 
is the {\it microscopic} elastic compatibility equation,
which relates strain modes:

$
(1-\cos k_x \cos k_y) e_1 (\vec{k})
- \sin k_x \sin k_y e_2 (\vec{k})
$

$
+ (\cos k_x - \cos k_y) e_3 (\vec{k}) = 0.  \hskip 33 mm  (1) 
$

\noindent The other two relate the intracell and the strain modes:

$
2 \cos \frac{k_x}{2} \cos \frac{k_y}{2} s_{\pm} (\vec{k})
\mp i \sin \left(\frac{k_x \pm k_y}{2} \right)
e_1 (\vec{k})
$

$
\pm i \sin \left(\frac{k_x \mp k_y}{2} \right) e_3 (\vec{k})
= 0.  \hskip 38 mm  (2)
$

\noindent These constraints
generate {\it anisotropic}
interactions (reflecting the lattice symmetry)
between atomic scale strain fields,
similar to the compatibility equations 
in Ref.~\cite{Shenoy99}, 
but now
including the intracell modes.
In the long wavelength limit, 
our description approaches the continuum model:  
For $\vec{k} \rightarrow 0$,
the above constraint equations 
can be written in real space as

$\nabla^2 e_1(\vec{r})
-2 \nabla_x \nabla_y  e_2 (\vec{r})
+ (\nabla_y^2-\nabla_x^2) e_3(\vec{r}) = 0, \hskip 7mm (3)
$

$
s_{\pm}(\vec{r})=
\left[ \left( \nabla_y  \pm
\nabla_x  \right) e_1(\vec{r})
+ \left(  \nabla_y \mp
\nabla_x \right) e_3(\vec{r}) \right]/4.  \hskip 5mm (4)
$

\noindent 
Equation~(3) is the usual compatibility equation 
in continuum theory.
Equation~(4) shows that 
{\it the spatial variations of strains
always generate intracell modes},
the magnitude of which vanish as the inverse 
of the length scale of the strain mode variations.
It is well-known in continuum Ginzburg-Landau theory that 
the energy associated with the 
gradient of strains
is responsible for domain wall energies as,
for example,
in structural phase transitions~\cite{Barsch84}.
The above result shows that the intracell modes are at the origin 
of such energy terms.
Since our strain-related variables
become identical to conventional strain variables
in the long wavelength limit,
various length scale lattice distortions may be described
in a {\it single} theoretical framework. 
This makes it possible to study typical multiscale 
situations where both short- and long-wavelength
distortions are important.
It also provides a natural framework for incorporating interactions
between atomic scale strain-related fields
coupled to 
other degrees of freedom in functional materials. 

The following analysis of 
the simple harmonic elastic energy for 
the square lattice
further exemplifies the utility of 
these variables.
We consider the simplest energy expression by
approximating
the total elastic energy by 
the sum of the elastic energy of each square:   
$$
E_{\text{sq.lat}}=\sum_{\vec{i}}\{ \sum_{n=1,2,3}
\frac{1}{2} A_n [e_n(\vec{i})]^2 + \sum_{m=+,-}
\frac{1}{2} B  [s_m(\vec{i})]^2 \},  (5)
$$

\noindent where $A_n$ and $B$ denote elastic moduli 
and `intracell modulus', respectively. The couplings between 
$e_1$, $e_2$, $e_3$, $s_+$, and $s_-$
at the {\it same} site are 
forbidden by symmetry
at the harmonic level, 
but are allowed at the anharmonic level,
which may have important consequences
for structural ``phase transitions'' at the nanoscale.
In Eq.~(5) 
the inter-atomic elastic energies 
between atoms beyond each square,
or further than the second nearest neighbors, 
are neglected.
These interactions may be included
by adding energy terms with distortion variables
at {\it different} sites, e.g., $e_1(\vec{i})e_1(\vec{i}+(10))$.
Since some of the atomic pairs are shared by two
square plaquettes of atoms,
the parameters in
Eq.~(5) should be 
appropriately renormalized. 
A robust way to determine the parameters is
to compare the phonon spectrum of our model
with experimental data.

\begin{figure}
\leavevmode
\epsfxsize8.5cm\epsfbox{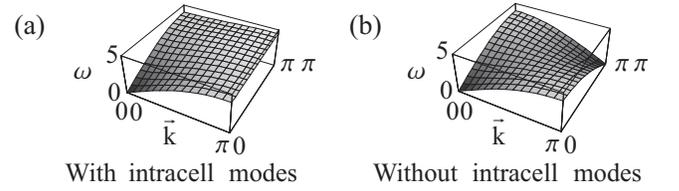}
\caption{\label{fig:phonon} 
An example of calculated phonon spectra
(a) with and (b) without 
intracell modes 
for a 2D square lattice
with a mono-atomic basis.
The upper phonon branch is shown for both cases
($M=\hbar=1$).
}
\end{figure}

For the lattice energy of Eq.~(5),
the phonon spectrum is given by
$\sqrt{M} \hbar \omega = \sqrt{E_1 \pm \sqrt{E_2}}$, 
where $E_1=( A_1 +A_2+A_3) (1-\cos k_x \cos k_y) /2 + B (1-\cos k_x) (1- \cos k_y)$,
$E_2=(A_1 +A_2-A_3)^2 \sin^2 k_x \sin^2 k_y /4 +
(A_1-A_2+A_3)^2 ( \cos k_x - \cos k_y)^2 /4$, 
and $M$ is the mass of an atom.
A typical 
spectrum (upper branch) for $A_1$=5, $A_2$=4, $A_3$=3, and $B$=5 
is shown in 
Fig.~\ref{fig:phonon}(a).
At $\vec{k}=(\pi,\pi)$, the distortion
is a pure intracell mode, and 
the energy depends only on 
the intracell mode modulus $B$.
Therefore, as shown in Fig.~\ref{fig:phonon}(b),
$\omega (\pi,\pi)$ vanishes
without the intracell mode ($B$=$0$),
which is unphysical.
As $k \rightarrow 0$,
the {\it slope} of the phonon spectrum 
is determined only by strain mode moduli, $A_1$, $A_2$, and $A_3$,
whereas the {\it curvature} depends on $B$ as well
because of Eq.~(4).
For $k \rightarrow 0$, 
since the intracell modes vanish as the inverse of the wavelength,
the lattice energy approaches
$ E_{\mathrm{sq.lat}}(\vec{k} \rightarrow 0) \approx
\sum_{\vec{k},n=1,2,3} A_n e_n^2(\vec{k}) / 2$,
in agreement with continuum theory.
Using Eq.~(4),
the energy for the intracell modes in Eq.~(5)
can be written as
$
E_{\text{intra}}\approx
\int d \vec{r}
B [  ( \vec{\nabla} e_1 )^2
+ ( \vec{\nabla} e_3 )^2 + 
2 ( \nabla_y e_1 \nabla_y e_3 -
\nabla_x e_1 \nabla_x e_3 ) ]/16.
$

We compare our approach to
a displacement-based Born-von K\'{a}rm\'{a}n
model~\cite{Kittel}
for the square lattice.
The first-nearest and the second-nearest neighbor 
atoms are connected 
by central-force and non-central-force
springs with spring constants $K_1^C$, $K_1^N$,
$K_2^C$  and $K_2^N$, respectively.
Elastic energies of the lattice, 
one from our model
and the other from the 
Born-von K\'{a}rm\'{a}n model, 
can be represented
in the following form
in terms of displacement variables, $d^x(\vec{i})$ and $d^y(\vec{i})$:
$E_{\text{sq.lat}}=\sum_{\vec{k},a,b} 
d^a(-\vec{k}) D_{ab}(\vec{k}) d^b(\vec{k})$. 
We find that the 
$D_{ab}(\vec{k})$'s for both models 
become 
identical if:  
$A_1=K_1^C-K_1^N + 2 (K_2^C-K_2^N)$, 
$A_2=2 (K_1^N + K_2^C + K_2^N )$,
$A_3 = K_1^C + K_1^N + 4 K_2^N$, and
$B = K_1^C+K_1^N$.

We apply our formalism to obtain the domain wall solution
for the atomic displacements between two homogeneous strain states
(a ``twin boundary'') due to a phase transition to a rectangular 
lattice.
We then compare the solution to that obtained from continuum theory
where discreteness effects are neglected~\cite {Barsch84}.
With elastic energy  $E_{\text{rec}}=E_{\text{rec}}^{(1)}+E_{\text{rec}}^{(2)}$, where

$
E_{\text{rec}}^{(1)}=\sum_{\vec{i}}
\frac{1}{2} A_1 e_1(\vec{i})^2 +
\frac{1}{2} A_2 e_2(\vec{i})^2 +
\frac{1}{2} B [ s_+(\vec{i})^2 + s_-(\vec{i})^2 ],
$

$
E_{\text{rec}}^{(2)}=\sum_{\vec{i}} -\frac{1}{2} A'_3 e_3(\vec{i})^2 +
\frac{1}{4} F_3 e_3(\vec{i})^4, \hskip 20 mm (6)
$

\noindent the degenerate ground state of $E_{\text{rec}}$ is a uniform state with 
$e_3$=$\pm \sqrt{A'_3/F_3}$, and
$e_1$=$e_2$=$s_+$=$s_-$=0.
To study the domain wall between these two degenerate 
rectangular ground states,
we consider $e_3(\vec{i})$ as the order parameter and
minimize $E_{\text{rec}}^{(1)}$ with respect to the other variables using 
the constraint equations [Eqs. (1) and (2)] and the method 
of Lagrange multipliers. 
We obtain $E_{\text{rec,min}}^{(1)}=\sum_{\vec{k}}
\frac{1}{2} e_3(-\vec{k}) U(\vec{k}) e_3(\vec{k})$, where
$U(\vec{k})=(V_1+V_2+V_3)/V_4$, and
$V_1= [ A_2 ( A_1 \beta_2^2 + A_2 \beta_1^2 )
+ B^2 \beta_4^2 ] A_1 \beta_3^2$,
$V_2=( 2 A_1 A_2 \beta_1^2 + A_2^2 \beta_1^2 + A_1^2 \beta_2^2 ) B \beta_1 \beta_4$,
$V_3=2(A_1 \beta_2^2 + A_2 \beta_1^2) B^2 \beta_4^2 +
B^3 \beta_1 \beta_4^3$,
$V_4=( A_1 \beta_2^2 +A_2 \beta_1^2
+ B \beta_1 \beta_4 )^2$,
with
$\beta_1=1-\cos k_x \cos k_y$,
$\beta_2=- \sin k_x \sin k_y$,
$\beta_3= \cos k_x - \cos k_y$, and
$\beta_4= (1-\cos k_x) (1-\cos k_y)$.

With $k_x$=$k \cos \theta $ and $k_y$=$k \sin \theta$,
the expansion of $U(k,\theta)$ about $k$=0 yields
$U(k,\theta)=U_{0}(\theta) + U_{2}(\theta)k^{2} + O(k^4)$, where 
$U_{0}(\theta) = A_1 A_2 \cos^2 2 \theta /(A_1 \sin^2 2\theta + A_2)$,
and
$U_{2}(\theta)=\sin^2 2 \theta [ 6 A_1 A_2 B \sin^2 2 \theta
+ 4 A_1 A_2 ( A_1 + A_2) \cos^2 2 \theta
+ 
3 B ( A_2^2 + A_1^2 \sin^2 2 \theta) ]/[ 24 (A_2 + A_1 \sin^2 2 \theta)^2 ] $.
\noindent 
The term $U_{o}$ is purely
orientation-dependent without a length scale, 
and is 
minimized at $\theta=$45$^o$ and 135$^o$,
as obtained in Ref.~\cite{Shenoy99}.
The difference between continuum and our discrete theory 
lies in the $k^2$ term:
continuum theory commonly assumes isotropic 
gradients in the order parameter, i.e., $(\vec{\nabla} e_3)^2$~\cite{Barsch84},
whereas $U_2(\theta)$ is anisotropic.
The two origins of the anisotropy are:
(a) the compatibility relation, Eq.~(1), 
which has higher powers in $k$ than Eq.~(3)
due to discreteness, 
and (b) the presence of
shuffle mode energy. 
The latter can be written as gradients
of strains, but with corrections to  
the phenomenological isotropic term, $(\vec{\nabla} e_3)^2$,
used in Ginzburg-Landau theory. 
As $U_2(\theta)$ is minimized for $\theta=0^o$ and $90^o$, 
it competes with $U_0(\theta)$  which prefers  
$\theta=45^o$ and $135^o$.
Thus, the domain wall direction depends on the length scale
with a critical length scale
$\lambda_c \sim \sqrt{B/A_1}$.
If $\lambda_c \le 1$, i.e., less than the interatomic spacing,
the domain wall has direction 45$^o$ or 135$^o$ 
down to atomic scales.
If $\lambda_c>1$, then for length scales smaller (larger)
than $\lambda_c$, the domain wall direction is 0$^o$ or 90$^o$
(45$^o$ or 135$^o$) and the domain wall has multiscale attributes.

\begin{figure}
\leavevmode
\epsfxsize8.5cm\epsfbox{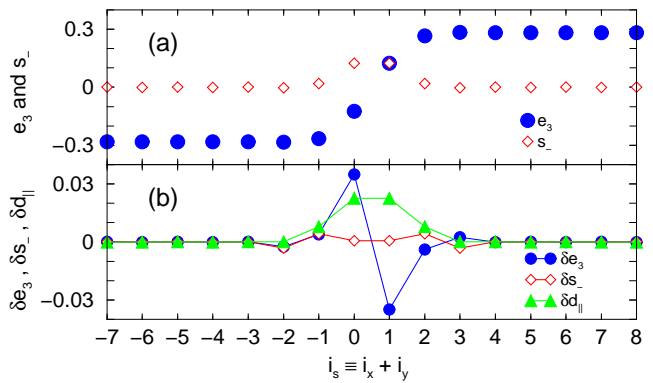}
\caption{\label{fig:wall.1d} (Color)
Atomic scale 135$^o$ domain wall profile for critical length scale,
$\lambda_c \le 1$ 
along the direction perpendicular to the domain wall:
(a) strain $e_{3}$ and shuffle $s_-$, 
(b) differences in $e_{3}$ 
($\delta e_{3}$=$e_{3,\mathrm{atomic}}-e_{3,\mathrm{continuum}}$), 
$s_-$ ($\delta s_{-}$) and
displacement parallel
to the domain wall direction ($\delta d_{||}$) between the results
from continuum theory
for ${\vec k} \sim 0$ and our model that 
includes discreteness.
The fields $e_1$, $e_2$, $s_+$, and displacement perpendicular
to the domain wall are zero.
Parameter values are $A_1=5$, $A_2=4$, $A'_3=4$, $B=5$, and $F_3=50$.
}
\end{figure}

\begin{figure}
\leavevmode
\epsfxsize8.5cm\epsfbox{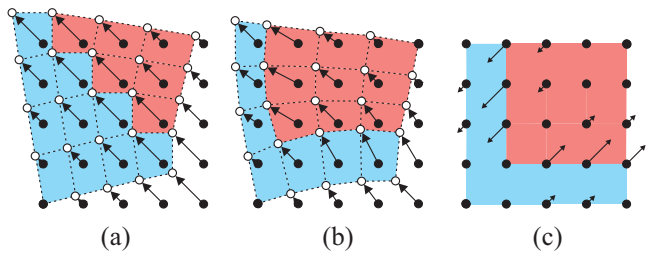}
\caption{\label{fig:disp123c} (Color)
Atomic displacements in the vicinity of domain wall:
(a) $\lambda_c \le 1$ and (b) $\lambda_c > 1$.
Color represents the sign of $e_3$ (red: positive, blue: negative),
and is lighter compared to 
Figs.~\ref{fig:wall.2d.neq} and ~\ref{fig:wall.2d.step}
to make the arrows visible.
(c) The displacement component 
perpendicular to the large scale domain wall 
direction for (b), magnified by a factor of 3.
Parameter values are $A_2=4$, $A'_3=4$, $B=5$, $F_3=50$, and
$A_1$ =5 for (a) and $A_1$=1 for (b) and (c).
}
\end{figure}

We  examine first the case $\lambda_c \le 1$ that would apply to 
materials with relatively large bulk modulus $A_1$ (`hard' materials)
for fixed $B$.
Here $k_x=\pm k_y$ and $U(\vec{k})=B(1-\cos k_x)/(1+\cos k_x)$.
The domain wall width is a result of the competition between 
$U(k)$ that favors $\vec{k} \rightarrow 0$, or thick domains,
and $E_{\text{rec}}^{(2)}$ that favors sharp walls. 
We illustrate the domain wall solution  with 135$^o$
domain wall direction. 
The only non-zero 
distortion modes
are $e_3$ and
$s_-$ ($s_+$ for a 45$^o$ domain wall).
The  strain $e_3$ reverses sign at the domain
wall, the intracell mode $s_-$ is confined within the domain wall,
and the atomic displacements are parallel to the domain wall direction. 
The numerical
solution~\cite{Euler}
 for $e_3$ and $s_-$ along a line perpendicular to the wall is shown
in Fig.~\ref{fig:wall.1d}(a), for which $\lambda_c \sim 1$. 
(Narrow domain walls with widths of a few unit cells, 
as considered here, 
have been identified experimentally~\cite{Chrosch99}.)
The corresponding displacement field 
near the center of the domain wall
is shown in Fig.~\ref{fig:disp123c}(a),
in which the red and blue colors  show regions with $e_3$
positive and negative, respectively.
Both figures show that the center of the domain wall is located at bonds 
rather than sites to avoid the higher energy state of $e_3$=0 and large $s_-$.
As for a Peierls-Nabarro barrier~\cite{PN}, the higher energy
(by $4.4 \times 10^{-4}$ per unit length for our parameter values)
for the site-centered domain wall  
acts as a pinning potential for the domain wall
due to the inherent discreteness. 
In Fig.~3(b) we  compare our results with continuum theory,
which predicts $e_3=e_3^{\text{max}} \tanh (i_s/\xi )$~\cite{Barsch84}
and $s_- = \partial e_3 /2 \partial i_s $ from Eq.~(4),
where $i_s=i_x+i_y$.
The differences in the interface region, 
shown in Fig.~\ref{fig:wall.1d}(b),
are of the order of $10\%$
of $e_3^{\text{max}}=\sqrt{A_3'/F_3}$. 
The domain wall width~\cite{Chrosch99}
is roughly given by $2\xi=\sqrt{2} \sqrt{B/A'_3}$
and the ratio between the maxima of $s_-$
and $e_3$, $s_-^{\text{max}}/e_3^{\text{max}}$, is about $1/(2\xi)$.

\begin{figure}
\leavevmode
\epsfxsize8.5cm\epsfbox{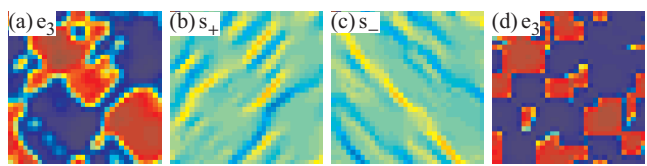}
\caption{\label{fig:wall.2d.neq} (Color)
Nonequilibrium domain wall state:
(a) $e_3$, (b) $s_+$, (c) $s_-$ for $\lambda_c \le 1$,
(d) $e_3$ for $\lambda_c>1$.
Parameter values are identical to Fig.~\ref{fig:disp123c}.
Dark red corresponds to 0.28 and dark blue to -0.28.
Green  implies a value close to zero.
}
\end{figure}

Anisotropic effects in $U(k,\theta)$ become more apparent
away from equilibrium, e.g., at finite temperatures or in other conditions
where metastability
is present.
Figure~\ref{fig:wall.2d.neq}(a) shows the results for $e_3$ of a
$2D$ simulation away from equilibrium~\cite{Noneq},
in which atomic scale domain walls are oriented along 45$^o$ and 135$^o$ directions.
The corresponding intracell modes are shown in Figs.~\ref{fig:wall.2d.neq}(b)
and~\ref{fig:wall.2d.neq}(c).
Note that both $s_+$ and $s_-$ shuffles are present at or near interfaces
and only one of these modes survives at equilibrium.
The horizontal or vertical ``jogs'' 
in the $s_{+}$ and $s_{-}$ walls
are secondary defects 
due to the competition between $U_0$ and $U_2$,
which provide principal
relaxation forces.

\begin{figure}
\leavevmode
\epsfxsize8.5cm\epsfbox{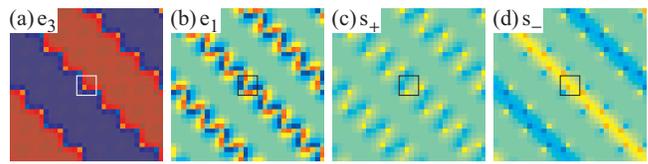}
\caption{\label{fig:wall.2d.step} (Color)
Atomic scale domain wall solution for materials with 
$\lambda_c>1$.
Parameters are the same as in Figs.~\ref{fig:disp123c}(b) and \ref{fig:disp123c}(c),
which show the region inside the square in this figure. 
Strain $e_2$ is zero. Color scheme is the 
same as in Fig.~\ref{fig:wall.2d.neq}.
}
\end{figure}

The domain wall solution for  $\lambda_c>1$, typical for small bulk modulus
$A_1$ or `soft' materials,
is shown in Fig.~\ref{fig:wall.2d.step}
for which $\lambda_c \sim \sqrt{5}$.
The $e_3$ field in Fig.~\ref{fig:wall.2d.step}(a) shows that on length scales
of the size of the system (larger than $\lambda_c$),
the diagonal orientation is still preferred.
However, this diagonal domain wall consists of a
`staircase' of $0^o$ and $90^o$ domain walls of length scale $\lambda_c$.
The existence of $0^o$ and $90^o$ walls in $e_3$ 
forces elastic compatibility to induce alternately large positive and negative
values in the dilatation strain $e_1$ in the horizontal and vertical parts
 of the `staircase', as shown in Fig.~\ref{fig:wall.2d.step}(b).
This has implications for the  functionality
of the domain walls. For example, the $e_1$ field can couple to charge and 
modulate the local charge density along the wall.
Similar features are also reflected in $s_+$ and $s_-$, as shown in 
Figs.~\ref{fig:wall.2d.step}(c) and \ref{fig:wall.2d.step}(d).
The displacement pattern within the square in Fig.~\ref{fig:wall.2d.step}
is shown in Fig.~\ref{fig:disp123c}(b).
Unlike the case  $\lambda_c \le 1$ in Fig.~\ref{fig:disp123c}(a),
the displacement has a component perpendicular
to the $135^o$ large scale domain wall direction,
which is shown in Fig.~\ref{fig:disp123c}(c) magnified by a factor of 3.
This component is greatest for the atoms  at the boundary
of the two domains.
A nonequilibrium state is shown in Fig.~\ref{fig:wall.2d.neq}(d).
The small domains have $0^o$ or $90^o$ boundaries,
but over larger length scales these domains are correlated
along $45^o$ and $135^o$ directions.

In summary, we have reported an approach
to ``atomic-scale elasticity'',
which uses symmetry modes of elementary objects
of atoms as distortion variables.
A gradient expansion for the energy with {\it anisotropic} coefficients 
has been obtained, with corrections to the usual 
phenomenological isotropic gradient terms used in 
Ginzburg-Landau theory.
As an illustration, we have
obtained   domain wall (twin boundary)
solutions in terms of strain {\it and} intracell modes 
and have shown
how they differ from the continuum elastic soliton solution~\cite{Barsch84}.
Our work provides the basis for interpreting
atomic scale features in HREM studies of 
domain walls~\cite{Stemmer95,Chrosch99}.

We thank A. J. Millis and S. R. Shenoy for insightful discussions.
This work was supported by the US DOE.


\begin{thebibliography}{99}
\bibitem{definition}
E.g.,
$e_3(i)=[-d^x(i)+d^y(i)+d^x(i+(10))+d^y(i+(10))
+d^x(i+(11))-d^y(i+(11))-d^x(i+(01))-d^y(i+(01))]/2\sqrt{2}$.
Care needs to be exercised in discretizing derivatives.
For example, $e_3(i)=\partial d^x / \partial x 
-\partial d^y / \partial y \approx
[ d^x(i+(10))-d^x(i) ]- [ d^y(i+(01))-d^y(i) ]$
does not reflect the symmetry of the square lattice.
\bibitem{Euler}
We use the Euler method(see, e.g., Ref.~\cite{Shenoy99})
with an appropriate boundary condition.
\bibitem{Noneq}
These states are found by probing the energy landscape with
the method of steepest descent,   
starting from  random initial conditions.
\end{thebibliography}
\end{document}